\documentstyle[aps,psfig,floats,epsf,twocolumn]{revtex}

\begin{document}
\tighten
\draft
\twocolumn[
\hsize\textwidth\columnwidth\hsize\csname @twocolumnfalse\endcsname  

\title{Raman scattering near a quantum critical point}
\author{J.K. Freericks$^*$ and T.P. Devereaux$^\dagger$}
\address{$^*$Department of Physics, Georgetown University, Washington, DC
20057, U.S.A.}
\address{$^\dagger$Department of Physics, University of Waterloo, Canada}
\date{\today}
\maketitle

\widetext
\begin{abstract}
Electronic Raman 
scattering experiments in a wide variety of materials (ranging from 
mixed-valence materials to Kondo 
insulators to high-temperature superconductors) show anomalous behavior in the
$B_{1g}$ channel when the system is on the insulating side of the 
metal-insulator transition.  Here we provide an exact solution for 
$B_{1g}$ Raman scattering in the Falicov-Kimball model, show how
these theoretical results are universal near the metal-insulator transition
and show how they produce the two main features seen in experiments near 
a quantum critical point: (i) the rapid appearance of low-energy spectral
weight as $T$ is increased from 0 and (ii) the existence of an isosbestic point
(where the Raman response is independent of $T$ at a characteristic frequency).
\end{abstract}
]
\narrowtext

Raman scattering has long been applied to study metals, insulators,
semiconductors, and superconductors. Via light's coupling to the electron's 
charge,
inelastic light scattering can reveal electron dynamics over a wide range
of energy scales and temperatures. More recently, Raman scattering
has played a strong role in elucidating the nature of electron
dynamics in the high temperature superconductors over the entire
region of their phase diagram. These studies have been particularly
useful in shedding light on superconducting and pseudogap energies
and on magnon scales and dispersions. Systems as disparate as
mixed-valence compounds (such as SmB$_{6}$ \cite{SLC1}), Kondo-insulators
(such as FeSi \cite{SLC2}), and the underdoped cuprate high temperature
superconductors \cite{irwin,hackl,uiuc}), 
show temperature-dependent $B_{1g}$
Raman spectra that are both remarkably similar and quite anomalous,
suggesting a common mechanism governing transport. As these
materials are cooled,
a pile up of spectral weight appears for moderate photon energy losses
with a simultaneous reduction of the low frequency spectral weight.
This spectral weight transfer is slow at high temperatures and then
rapidly increases as temperature is lowered towards a putative
quantum critical point (corresponding to a metal-insulator transition). 
In addition, the spectral range
is divided into two regions: one where the Raman response decreases as $T$
is lowered and one where the response increases.  These regions are
separated by a so-called isosbestic point, which is defined to be the
characteristic frequency where the Raman response is independent of
temperature.

While the existence of anomalous features in Raman scattering near
a quantum critical point has been seen for some time,
there is no theoretical understanding
that connects the metallic and insulating states.
In 1991, Shraiman and Shastry\cite{ss} outlined a procedure to construct 
a theory that can interpolate from a metal to an insulator. However, due to 
the ever increasing Hilbert space needed
to describe a metal from the insulating side, or the lack of a
clear picture of quasiparticles from the metallic side, quantitative
calculations were not feasible. One is then left with approximate methods (such
as perturbation theory for the Hubbard model) or more phenomenological
approaches to construct a Raman theory which suffer the limitations of
being unable to reach different phases and of lacking a microscopic basis.

Here we provide the first exact theoretical description of Raman scattering
near a quantum metal-insulator transition that contains all of the
anomalous behavior seen in experiments on these strongly correlated materials.
We choose the spinless Falicov-Kimball model\cite{falicov_kimball}
as our canonical model for Raman scattering. It
contains two types of electrons: itinerant band electrons and localized
(d or f) electrons.  The band electrons can hop between nearest neighbors
[with hopping integral $t^*/(2\sqrt{d})$ on a $d$-dimensional cubic
lattice],
and they interact via a screened Coulomb interaction with the localized
electrons (that is described by an interaction strength $U$ between
electrons that are located at the same lattice site). We measure all
energies in units of $t^*$.
The Hamiltonian is
\begin{eqnarray}
H =-\frac{t^*}{2\sqrt{d}}\sum_{\langle i,j\rangle}d_i^{\dagger}d_j +&&E_f
\sum_i w_i-\mu\sum_i(d_i^{\dagger}d_i+w_i)\nonumber\\
&&+U\sum_id_i^{\dagger}d_iw_i,
\label{eq: ham}
\end{eqnarray}
where $d_i^{\dagger}$ $(d_i)$ is the spinless conduction electron creation
(annihilation) operator at lattice site $i$ and $w_i=0$ or 1 is a classical
variable corresponding to the localized $f$-electron number at site $i$. 
We will adjust
both $E_f$ and $\mu$ so that the average filling of the $d$-electrons is 1/2 
and the average filling of the $f$-electrons is 1/2 ($\mu=U/2$ and $E_f=0$).

\begin{figure}
\vspace{5mm}
\centerline{\psfig{file=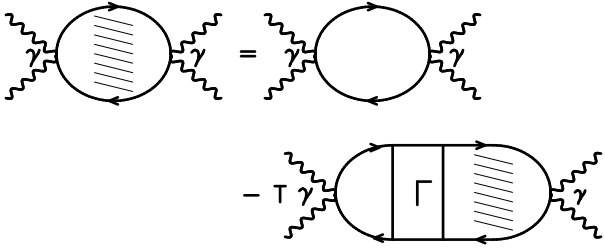,width=8cm,silent=}}
\vspace{.5cm}
\caption[]{Dyson equation for the nonresonant Raman response function.
Solid lines denote electron propagators and wavy lines denote
photon propagators. The shading denotes the fully renormalized susceptibility
and the symbol $\Gamma$ is the irreducible frequency-dependent charge vertex.}

\end{figure}                                                                   

The Raman response is found from the frequency-dependent
density-density correlation function that is depicted in Figure 1. 
The Raman scattering process involves a two-photon-electron-hole vertex
function that is called the Raman scattering amplitude $\gamma({\bf k})$.
In addition to the effects of the
Raman scattering amplitude, the density-density
correlation function is, in general, also
renormalized by the irreducible dynamical
charge vertex, which is denoted by $\Gamma$ in Figure 1. Evaluating the
diagrams in the standard fashion produces the following result:
\begin{eqnarray}
&&\chi(i\nu_l)=\sum_{\bf k}\int_0^{\beta}d\tau e^{i\nu_l\tau}\\
&&\times
\Biggr\{ \frac{{\rm Tr}T_\tau\langle e^{-\beta H}\rho_{\bf k}(\tau)\rho_{\bf k}
(0)
\rangle}{Z}-\Biggr [ \frac{{\rm Tr}\langle e^{-\beta H}\rho_{\bf k}(0)\rangle}
{Z} \Biggr ]^2\Biggr\},
\label{eq: raman}
\nonumber
\end{eqnarray}
with the uniform (${\bf q}=0$) Raman density operator
\begin{equation}
\rho_{\bf k} = \gamma({\bf k})d^{\dagger}_{\bf k}d_{\bf k}, 
\quad d_{\bf k}=\frac{1}{N}\sum_j e^{-{\bf R}_j\cdot {\bf k}}d_j,
\label{eq: ramandensity}
\end{equation}
$Z=Tr\langle e^{-\beta H}\rangle$, the partition function, 
and $i\nu_l=2i\pi lT$ the bosonic Matsubara frequency (the $\tau$-dependence
of the operators is with respect to the full Hamiltonian).
The Raman scattering amplitude $\gamma({\bf k})$ is a complicated function of 
the incoming and outgoing photon polarizations, of the photon energies, and the
polarizability of the medium.  In nonresonant Raman scattering 
one neglects the frequency dependence of the Raman scattering
amplitude,
and characterizes the Raman response in terms of the different spatial 
symmetries of the remaining function $\gamma({\bf k})$.  One can expand
this function in a Fourier series and examine the contributions of the lowest
components of the series, and compare them to experiment (the lowest order
$B_{1g}$ contribution is $\gamma({\bf k})=\sum_{j=1}^d(-1)^j\cos {\bf k}_j$,
with $d\rightarrow\infty$ the spatial dimension).  
More sophisticated
approaches would calculate the Raman scattering amplitude
from ``first-principles'' and
would include any possible resonant Raman scattering effects.  We leave those
pursuits to future work.

The Falicov-Kimball model can be solved exactly in the infinite-dimensional
limit by using dynamical mean-field theory (see Ref. \cite{bf} for details). 
The dynamical
charge vertex is local in infinite dimensions which implies that correlation
functions that have the same symmetry as the lattice are renormalized due
to this charge vertex, but correlation functions that are orthogonal
to the lattice, have no vertex corrections, and so they are represented
by their bare bubble diagrams\cite{khurana}. Although the fully symmetric
$A_{1g}$ Raman response can be calculated directly, we concentrate here
on the $B_{1g}$ response, which is described by the bare bubble diagram
in Figure 1. In this case, a straightforward symmetry analysis also shows that
resonant Raman scattering effects vanish for
nearest neighbor hopping on a hypercubic lattice in large dimensions, so the
total Raman response is represented by these nonresonant results. 

The Falicov-Kimball model has a ground state that is not a Fermi liquid because
the lifetime of a quasiparticle is finite at the Fermi energy.  As $U$ 
increases, the system first enters a pseudogap phase, where spectral weight
is depleted near the chemical potential, and then undergoes a metal-insulator
transition.  The interacting density of states (DOS)
is, however, temperature-independent for fixed $U$ and fixed electron 
fillings. For half-filling, $U<0.65$ corresponds to a weakly-correlated
metal, while a pseudogap phase appears for $0.65<U<1.5$ moving through
a quantum critical point at $U=1.5$ to the insulator phase $U>1.5$ (we neglect
all possible charge-density-wave phases here). 

\begin{figure}
\vspace{5mm}
\centerline{\psfig{file=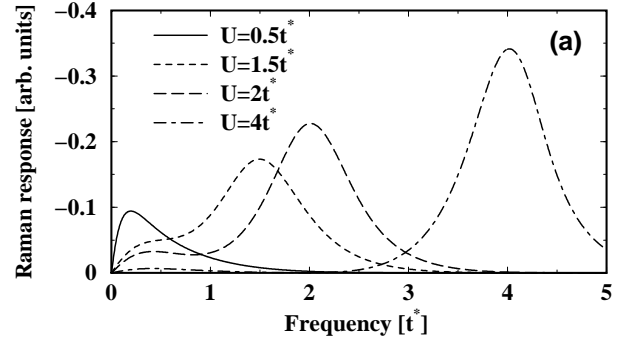,width=8cm,silent=}}
\vspace{.5cm}
\caption[]{$B_{1g}$ Raman response as a function of $U$ at $T=0.5$.
}
\label{fig2}
\end{figure}   

The Raman response is calculated in a similar fashion to the dynamical
charge susceptibility\cite{shvaika}: the Dyson equation
of Figure 1 is formally analytically continued to the real axis, where
the expression for the Raman response is found to depend solely on
complex integrals of the interacting single-particle DOS.
In Figure \ref{fig2} we plot the Raman response at a fixed temperature
$T=0.5$ for different values of $U$. For small values of $U$,
a small scattering intensity is observed due to the weak interaction
among ``quasiparticles'' providing a small region of phase space allowable for
pair scattering. The peak of the response reflects the dominant
energy scale for scattering, as is well known in metals\cite{metals}
and the high-energy tail is the cutoff determined by the finite energy band.
As $U$ increases, the low-frequency response is depleted as
spectral weight gets shifted into a large charge transfer peak at
a frequency $\sim U$. The charge transfer peak begins to appear for
values of $U$ for which the DOS is still finite at the Fermi level
and becomes large in this pseudogap phase before growing even larger in
the insulating phase. Notice how low-frequency spectral weight remains
even as one is well on the insulating side of the quantum critical point.  It
is these spectral features that are characteristically seen in the experiments
and which can only be seen in a theory that approaches the quantum critical
point.     
 
\begin{figure}
\vspace{5mm}
\centerline{\psfig{file=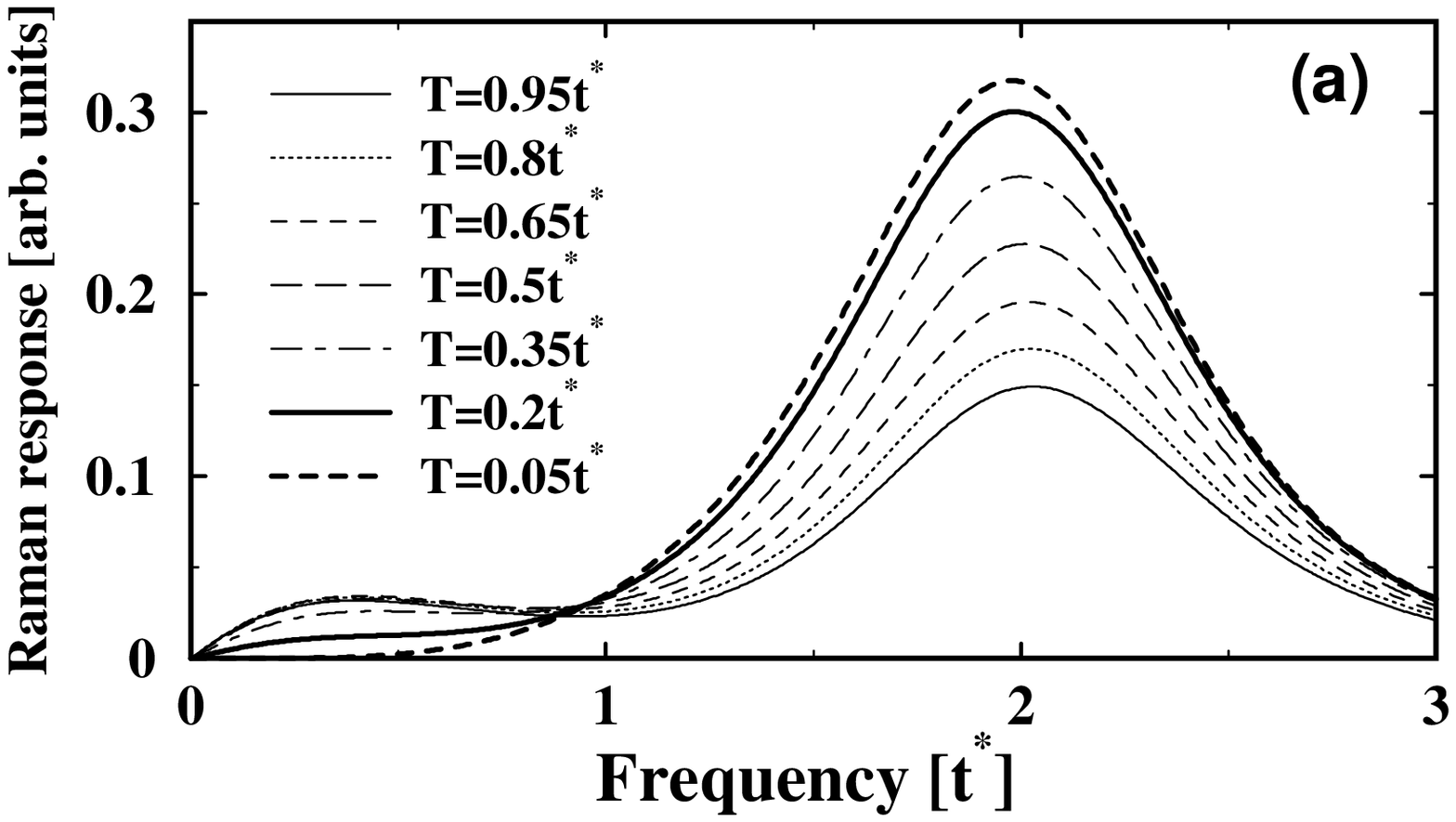,width=7.6cm,silent=}}
\centerline{\psfig{file=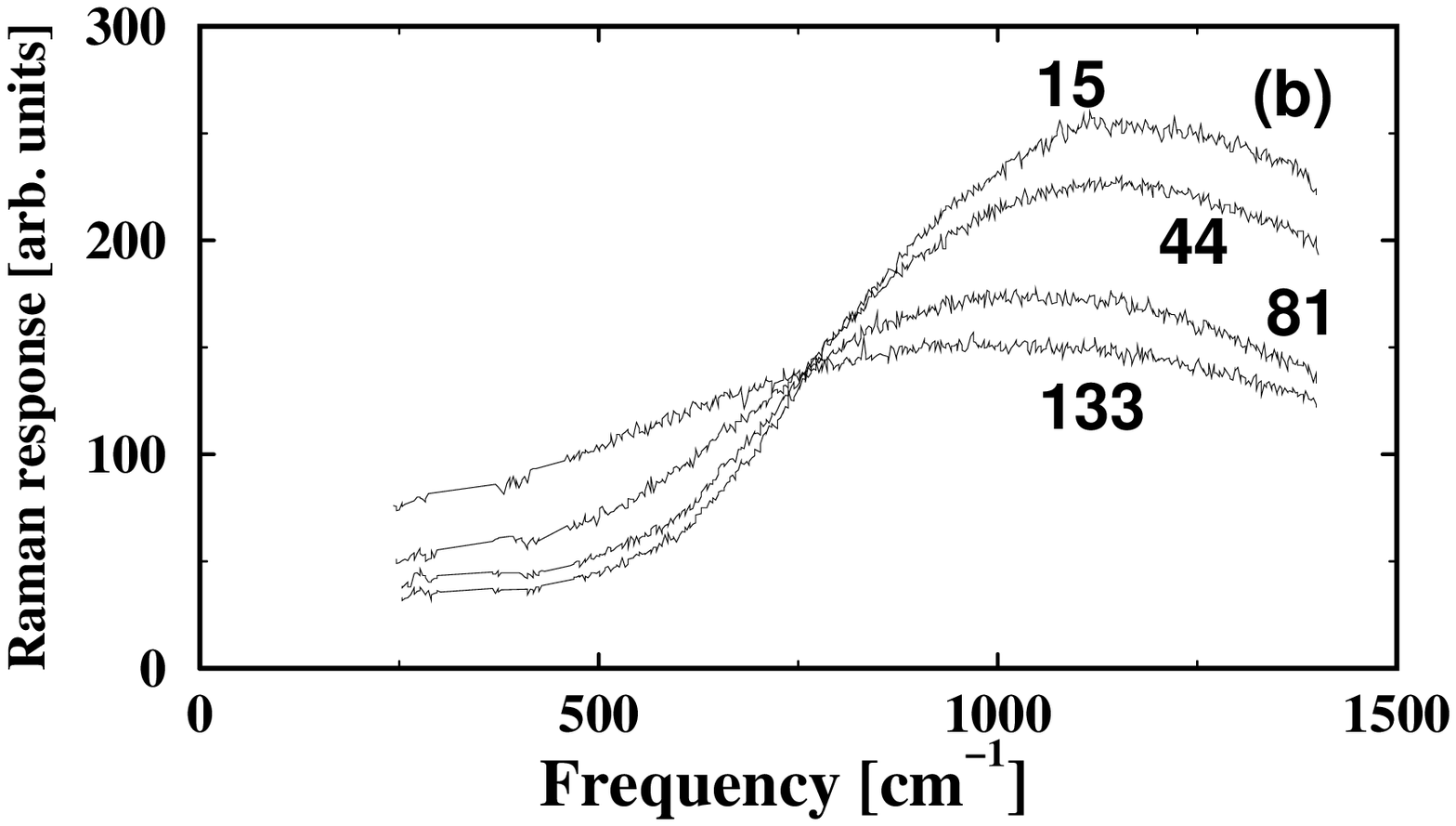,width=8cm,silent=}}
\vspace{.5cm}
\caption[]{$B_{1g}$ (a) Theoretical
Raman response as a function of temperature for $U=2$
(which lies just on the insulating side of the metal-insulator transition) and
(b) experimental\cite{SLC2}
Raman response for FeSi at moderate to low temperatures
(where the isosbestic point and low temperature spectral weight depletion
is seen). The experimental graphs are labeled by the temperature (in K)
where the data was collected.
}
\label{fig3}
\end{figure}   

In Figure \ref{fig3}(a) we plot the temperature dependence on the insulating
side of the metal-insulator transition. For small values of
$U$ the spectra are only slightly dependent on temperature due largely to
the small changes in kinetic energy with $T$.
As $U$ increases into the pseudogap and insulating phases, nontrivial
temperature
dependences begin to appear. The total spectral weight increases dramatically
with decreasing temperature as charge transfer processes become
more sharply defined. At the same time, the low-frequency response
depletes with lowering temperatures, vanishing at a temperature which is
on the order of the $T=0$ insulating gap. This behavior is precisely what
is seen in experiment\cite{SLC2} on FeSi at low temperatures \ref{fig3}(b)
where both
the isosbestic point and the low temperature spectral weight depletion
can be seen. 

We attribute the presence of a low-frequency response in a system which
is a strongly correlated insulator
to the appearance of thermally activated transport channels.
In the insulating phase at zero temperature, the only available intermediate
states created by the light must involve double site occupancy of a conduction
and a localized electron, with an energy cost of $U$. This gives the large
charge transfer peak at an energy $U$. As the temperature is increased,
for half filling, double occupancy can occur and as a result light can
scatter electrons to hop between adjacent unoccupied states either directly
or via virtual double occupancies. The number of electrons
which can scatter in this fashion increases with increasing temperature,
leading to an increase in the low-frequency spectral weight. The frequency
range for this low-frequency Raman response is determined by the lower
Hubbard bandwidth, which is typically much larger than the temperature
at which these features first appear. If one were to interpret
the temperature at which the Raman spectral weight starts to deplete as the
transition temperature
$T_c$ and the range of frequency over which the weight is depleted as the gap
$\Delta$, then one would conclude that near the quantum critical point $2\Delta/
k_BT_c \gg 1$. This is because the ``$T_c$'' is determined by the gap
in the single-particle density of states (which approaches zero at the
quantum critical point), while the ``$\Delta$'' is
determined by the width of the lower Hubbard band (which remains finite at
the quantum critical point); hence the ratio can become
very large near the quantum critical point.

The spectral weight transfer from low frequencies to the charge
transfer peak as a function of temperature can be quantified by
separating the Raman response into two regions determined by the isosbestic
point and plotting the 
total low-frequency spectral weight versus temperature (not shown). 
Choosing $U/2$ as the location of the isosbestic point, we find
that the reduction of spectral weight from high to low
temperatures is over 50 percent even in the weak pseudogap phase, and 
decreases by well over three orders
of magnitude as $U$ increases into the insulating phase ($U=4$).

\begin{figure}[htb]
\vspace{5mm}
\centerline{\psfig{file=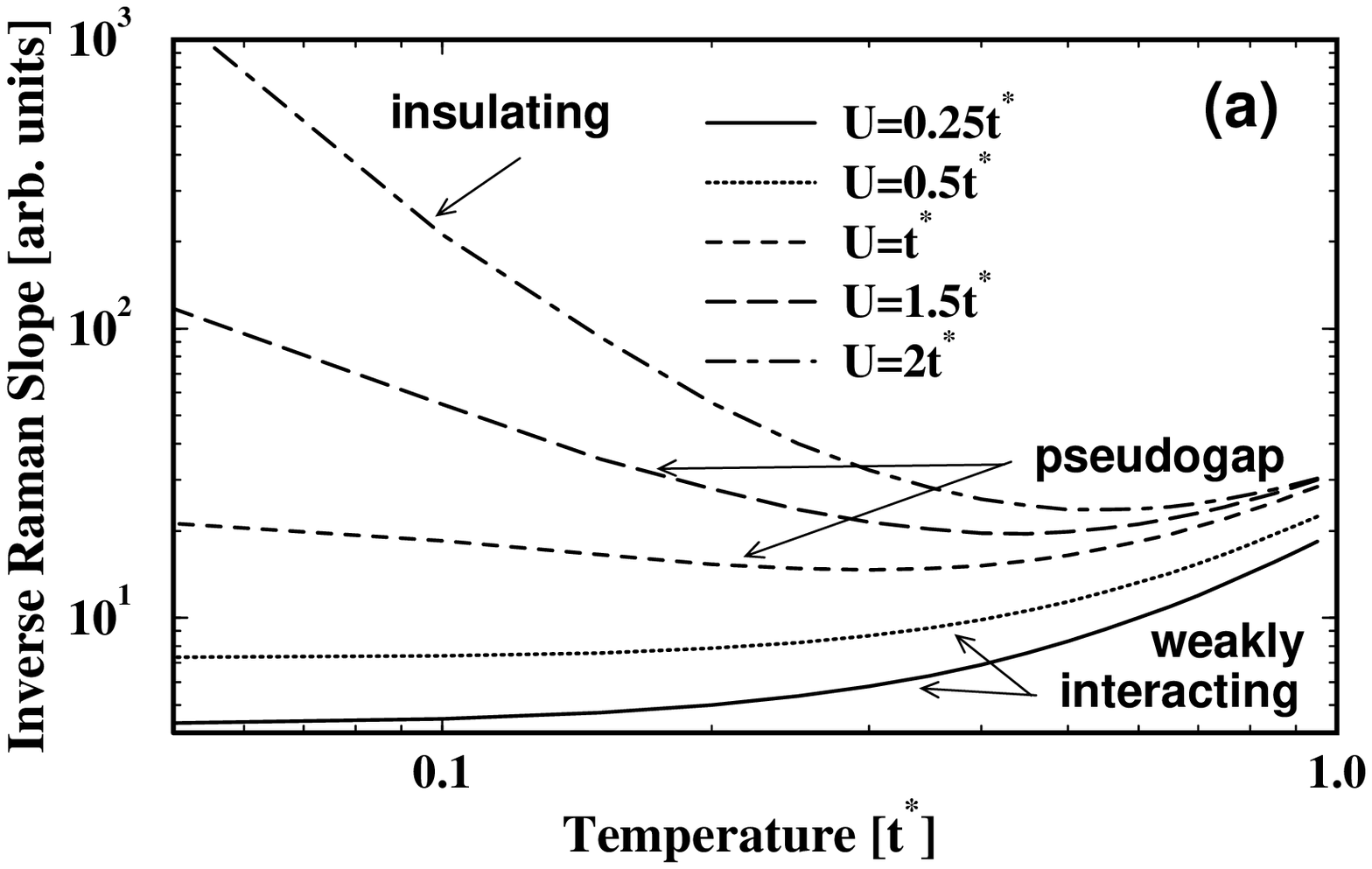,width=8.0cm,silent=}}
\centerline{\psfig{file=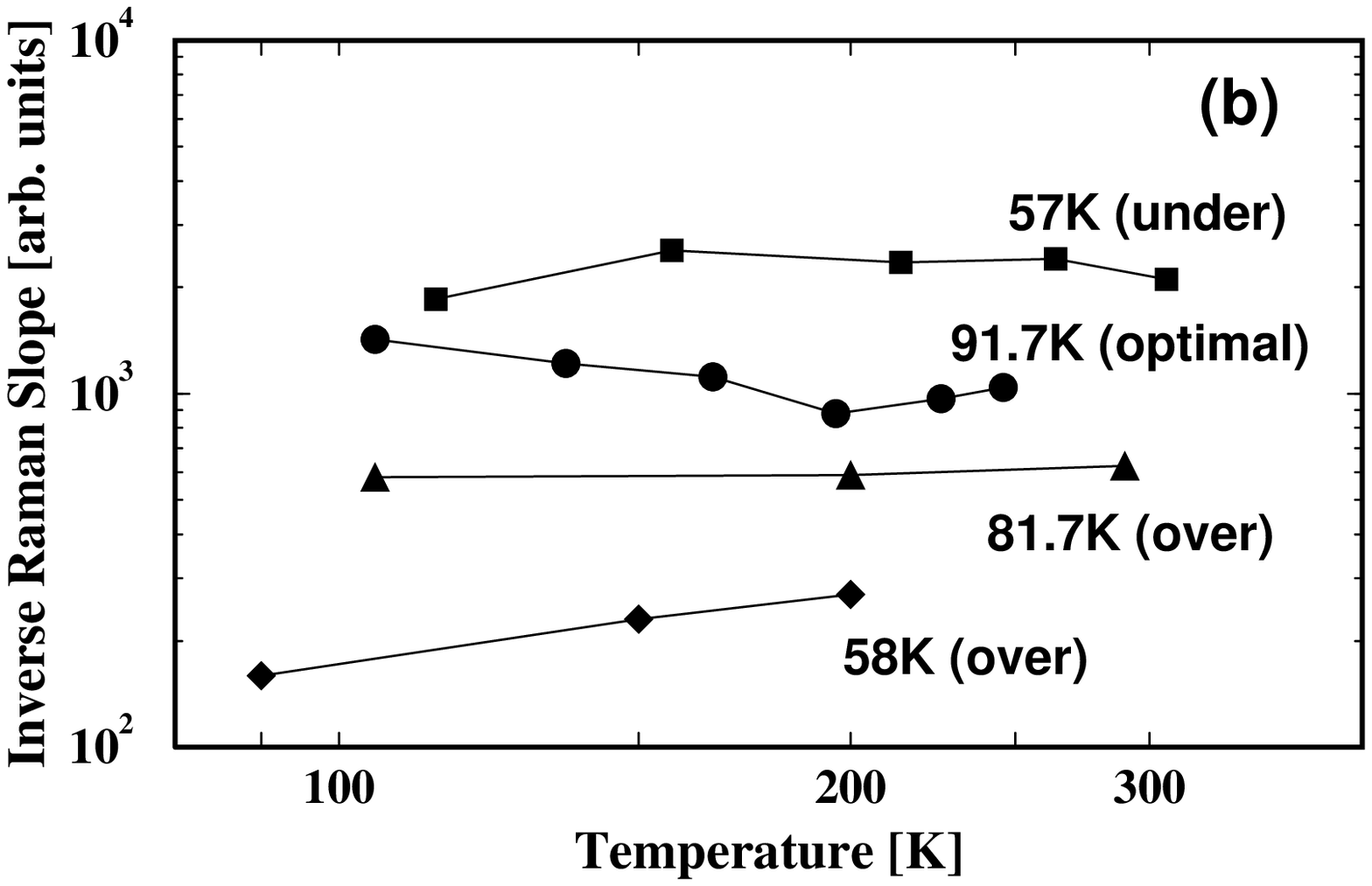,width=7.7cm,silent=}}
\vspace{.5cm}
\caption[]{(a) Theoretical inverse slope of the $B_{1g}$ Raman response and
(b) experimental\cite{irwin} inverse slope for optimal and 
underdoped high-temperature superconductors (the curves are labeled by their
superconducting transition temperature and whether they are under or 
overdoped; the 57K and 81.7K data have been multiplied by constant factors to 
separate them).  }
\label{fig5}
\end{figure}  

Lastly, we plot the inverse slope of the Raman response in Figure 
\ref{fig5}(a) as a function of temperature for different values of $U$
and the experimental\cite{irwin}
results on optimal and underdoped copper oxides \ref{fig5}(b).
Since the self energy is
temperature independent, we might expect a constant Raman
slope as a function of temperature, as is the case
with disordered noninteracting electrons. However, this is not the
case due to the formation of a thermally generated band
for scattering.  For small values of $U$, the
temperature dependence of the Raman
inverse slope is weak due to the temperature independence of the
self energy. However, as the single-particle bands begin to separate,
the relevance of thermally generated
double occupancies becomes more pronounced and the inverse slope
becomes temperature dependent at low temperatures. We see that
as $U$ increases the low temperature slope increases dramatically
due to the depletion of low-frequency spectral weight.
In particular, even in the pseudogap phase the inverse slope rises with
decreasing temperature indicating the proximity to the quantum-critical
metal-insulator transition. As $U$ increases into the insulating phase,
the temperature dependence of the Raman inverse slope is indicative
of the formation of gapped excitations.  Note how similar the experimental
results are to the theoretical predictions (the lowest temperature data for the
underdoped case (top curve) has large error bars because the signal is so
small; in all cases as the system becomes underdoped the $B_{1g}$ response
behaves more insulator-like).

In more complicated correlated models of the metal-insulator transition
the single-particle density of states will have a Fermi liquid peak at
low frequencies which may add new features to the Raman response, but on
the insulating side of the transition, where most of the anomalous
behavior is seen, the single-particle density of states should be
very similar to that of the Falicov-Kimball model (except for some additional
weak temperature dependence of the interacting DOS), which is why these
results are generically expected to be model-independent. 

Our theoretical results compare quite favorably to the experimental results
seen in a wide range of different materials ranging from mixed-valence
compounds \cite{SLC1}, to Kondo insulators \cite{SLC2} to the 
underdoped high-temperature 
superconducting oxides \cite{irwin,hackl,uiuc}.  In particular, all of those 
experimental systems appear to be close
to, but on the insulating side of the metal-insulator transition, and hence
they illustrate the two characteristic behaviors seen in our theory: (i) there
is a rapid rise in the low-frequency spectral weight at low temperatures (at
the expense of the high-frequency spectral weight) and
(ii) there is an isosbestic point.  Our model always produces an isotropic gap,
so we are unable to illustrate some of the behavior seen in the copper-oxides
where only the $B_{1g}$ response is anomalous, and the $A_{1g}$ and $B_{2g}$
responses are metallic rather than insulating.  But our results do indicate
a ``universality'' and model independence of the Raman response on the 
insulating side of, but in close proximity to, a quantum critical point.
We believe this is the reason why so many different materials show the
same generic behavior in their electronic Raman scattering.

\acknowledgments

J.K.F. acknowledges support of the National Science Foundation under grant 
DMR-9973225.  T.P.D. acknowledges support from the National Research and
Engineering Council of Canada. 
We also acknowledge useful discussions with S.L. Cooper, R. Hackl,
J.C. Irwin, M.V. Klein, P. Miller and A. Shvaika. We also thank S.L. Cooper,
R. Hackl, and J.C. Irwin for allowing us to reproduce their data.

\end{document}